\documentclass[prb,twocolumn]{revtex4}%
\usepackage{times}
\usepackage{graphicx}
\usepackage{dcolumn}
\usepackage{bm}
\usepackage[usenames,dvipsnames]{color}
\usepackage{amsmath}
\usepackage{amsfonts}
\usepackage{amsthm}
\usepackage{amssymb}
\usepackage{mathrsfs}
\usepackage{subfigure}
\usepackage{ulem}
\usepackage{verbatim}
\usepackage{hyperref}

\begin{document}

\title{Influence of structural deformations on the reentrant conductance feature in semiconducting nanowires}
\author{Iann Cunha}
\author{Leonardo Villegas-Lelovsky}

\author{Leonardo Kleber Castelano}
\email{lkcastelano@ufscar.br}
\affiliation{Departamento de F\'{\i}sica, Universidade Federal de S\~ao Carlos, 13565-905, S\~ao Carlos, SP, Brazil}
\affiliation{Departamento de F\'{\i}sica, IGCE, Universidade Estadual Paulista, 13506-900 Rio Claro SP, Brazil}

\date{\today}
\begin{abstract} % abstract
 Helical states can be measured through the observation of the reentrant behaviour, which is a dip in the conductance probed in semiconducting nanowires (NWs) with strong spin-orbit coupling (SOC) under the presence of an external perpendicular magnetic field. We investigate the effects of deformations in the electronic transport in NWs considering the coupling between different transverse modes. Within this approach, we show that the dip in the conductance of a NW is affected by the presence of a local constriction. Moreover, we find that the reentrant feature in the conductance can appear in NWs with a local expansion of its radius, even in the absence of SOC and magnetic field. Furthermore, we develop a numerical approach to calculate transport properties, which is able to include the deformation and the coupling among several scattering channels.
\end{abstract}

 \maketitle
 \section{Introduction}
 Majorana zero modes can be employed to perform topologically protected quantum computation ~\cite{Nayak}. The observation of Majorana states was predicted to occur in a system composed of a semiconducting nanowire (NW) with strong spin-orbit coupling (SOC) in proximity to an s-wave superconductor and under the presence of an external magnetic field~\cite{Alicea,Oreg,Lutchyn}. The experimental realization and characterization of semiconductor NWs based on InAs and InSb have been explored to check the existence of helical states~\cite{Heedt,Kammhuber,Weperen,Sun}, which are related to Majorana zero modes. The verification of helical states can be probed by the so-called reentrant feature, which appears as a measurable dip in the conductance. The reentrant feature was predicted to be observed when the electron energy reaches the pseudo gap occurring in NWs with strong Rashba SOC and subjected to an external magnetic field~\cite{Streda}. In our approach~\cite{our}, we reinterpret the dip in the conductance as a resonant reflection, which occurs if the NW contains a localized attractive potential and the coupling between different scattering channels is different from zero~\cite{Gurvitz}. Furthermore, Sanchez {\it et al}~\cite{Sanchez} showed that the Rashba Hamiltonian limited to a finite region is equivalent to an attractive potential and promotes the coupling of different transport channels, when a gauge transformation is taken into account. Therefore, the dip in the conductance can exist even in the absence of the magnetic field~\cite{Heedt,our}.  In this paper, we show that these fundamental ingredients can also be found in NWs having only a local structural deformation. The study of conductance considering structural deformations has a long history and several interesting phenomena have been found over the past decades\cite{PhysRevB.49.16581,PhysRevB.53.4054,PhysRevB.60.10962,PhysRevB.60.3963,PhysRevB.71.115430,PhysRevB.71.155330,PhysRevB.79.155305,Wo_oszyn_2014,SHCHAMKHALOVA200551}. Perhaps, the most important one, was the discovery of quantum point contacts\cite{qpc,qpc1} which are narrow constrictions in a two dimensional electron gas that exhibit quantized conductance and can be used as sensitive charge detectors.
  As already mentioned, here we show that the appearance of the dip in the conductance can be observed even when the Rashba SOC is lacking, but the NW must have a limited region of space where its radius suffers an expansion. On the other hand, a local radius compression does not trigger the dips in the conductance because the fundamental ingredient of having an attractive potential is lacking in this case. We also probe the NW with structural deformations and the Rashba SOC occurring in the same region of space. Moreover, we consider a perpendicular magnetic field in the whole NW together with Rashba SOC and structural deformations. Results, in this case, show that structural deformations can affect the observation of the reentrant feature and even mitigate it. Lastly, we provide a scheme to numerically solve the transport of charges considering the coupling among several scattering channels and the spatial dependence of the localized potential, which made the realization of the study of structural deformations in NW possible. This method is very useful and can be employed to investigate other systems, {\it e.g.} to study the transport of holes with different effective masses coupled between each other in a heterostructure using the Luttinger Hamiltonian~\cite{Winkler}.
 %%%%%%%%%%%%%%%%%%%%%%%%%%%%%%%%%%%%%%%%%%%%%%%%%%%%%

\section{Theoretical Model}
Our model to study the electronic transport in an NW with a structural deformation is described by the following Hamiltonian.
\begin{equation}
\left[\frac{-\hbar^2}{2m^*}\nabla^2+V_c(x,y)+V(x,y,z)\right]\Psi(x,y,z)=E\Psi(x,y,z),\label{hamiltonian}
\end{equation}
where $m^*$ is the effective mass, $V_c(x,y)$ is the lateral confining potential, and $V(x,y,z)$ is the potential describing the structural deformation. To solve (\ref{hamiltonian}), we use the transverse modes $\Phi_{n,m}(x,y)$, which are solutions of the following equation
\begin{equation}
\left[\frac{-\hbar^2\nabla^2_\perp}{2m^*}+V_c(x,y)-E_{n,m}\right]\Phi_{n,m}(x,y)=0,\label{hamiltoniant}
\end{equation}
where $\nabla^2_\perp=\left(\frac{\partial^2}{\partial x^2}+\frac{\partial^2}{\partial y^2}\right)$.
The wave-function can be expanded in the complete basis formed by the transverse modes (channels) such as $\Psi(x,y,z)=\sum_{n,m}\chi_{n,m}(z)\Phi_{n,m}(x,y)$, where $\chi_{n,m}(z)$ are the scattering wave-functions, which are solutions of the coupled-channels equations
\begin{equation}
\left[\frac{\partial^2}{\partial z^2}+k^2_{n,m}\right]\chi_{n,m}(z)-\frac{2m^*}{\hbar^2}\sum_{n',m'}V^{n,m}_{n',m'}(z)\chi_{n',m'}(z)=0, \label{hamiltonianchannel}
\end{equation}
where $k^2_{n,m}=\frac{2m^*}{\hbar^2}(E-E_{n,m})$ and
\begin{equation}
V^{n,m}_{n',m'}(z)=\int dx\int dy\; \Phi^*_{n,m}(x,y)V(x,y,z)\Phi_{n',m'}(x,y).
\end{equation}
To be able to solve (\ref{hamiltonianchannel}), we assume that $V(x,y,z)$ is different from zero only in the region $|z|\leq L/2$. Therefore, there are three distinct regions where we must calculate the wave-functions. Usually, wave-functions are written in a vectorial form, but here we must use the matrix form to be able to implement the numerical scheme. This matrix form also has the advantage of considering the injection of electrons in different channels and solving these different channels at once. In the region I, the matrix elements of the wave-function $\chi_{\text{I}}$ are given by
\begin{eqnarray}
    \chi^{\text{I}}_{i,j}= 
\begin{cases}
    e^{i z k_i}+r_{i,i}e^{-i z k_i} ,& \text{if } i=j\\
    r_{i,j}e^{-i z k_i} ,              & \text{if } i\neq j.
\end{cases}\label{chi1}
\end{eqnarray}
In regions II and III, the matrix form for the wave-function respectively are
\begin{eqnarray}
\label{chi2}
\chi^{\text{II}}=
\left(
\begin{array}{ccccc}
 \phi _{1,1}(z)& \phi _{1,2}(z)& \ldots& \phi _{1,N}(z)  \\
 \phi _{2,1}(z)& \phi _{2,2}(z)& \ldots& \phi _{2,N}(z)  \\
  \phi _{3,1}(z)& \phi _{3,2}(z)& \ldots& \phi _{3,N}(z)  \\
  \vdots&\vdots&\ddots&\vdots\\
 \phi _{N,1}(z)& \phi _{N,2}(z)& \ldots& \phi _{N,N}(z)  \\
\end{array}
\right)
,
\end{eqnarray}

\begin{eqnarray}
\label{chi3}
\chi^{\text{III}}=
\left(
\begin{array}{ccccc}
t_{1,1}e^{i z k_1}& t_{1,2}e^{i z k_1}& \ldots & t_{1,N}e^{i z k_1}\\
t_{2,1}e^{i z k_2}&t_{2,2}e^{i z k_2}& \ldots & t_{2,N}e^{i z k_2}\\
t_{3,1}e^{i z k_3}& t_{3,2}e^{i z k_3}& \ldots & t_{3,N}e^{i z k_3}\\
\vdots&\vdots&\ddots&\vdots\\
t_{N,1}e^{i z k_N}& t_{N,2}e^{i z k_N}& \ldots &t_{N,N}e^{i z k_N}
\end{array}
\right).
\end{eqnarray}
The wave-function $\chi^{R}$ is the solution of (\ref{hamiltonianchannel}) in the region $R$, where $R=$I corresponds to $z\leq-L$/2, $R=$II corresponds to $|z|\leq L/2$, and $R=$III corresponds to $z\geq L/2$. We also assume that electrons are injected from the left ($z=-\infty$) in the channel $i$. The channel labeled by $i$ corresponds to the quantum numbers $(n,m,s)$ in (\ref{hamiltonianchannel}) in ascending order of the energy of transverse modes $E_{n,m}$, where $s=\pm$ designates the spin degree of freedom. Particularly, the Hamiltonian (\ref{hamiltonian}) does not depend on spin and transport equations can be solved to each spin separately. Although there are an infinite number of transverse modes, we truncate the problem to a certain size $N$. The coefficients $r_{i,j}$ and $t_{i,j}$ represent the reflection and transmission coefficients of an electron injected in $j$-th channel and scattered in $i$-th channel. The $\phi_{i,j}(z)$ are the wave-functions in the scattering region and depend on the potential form, which can be rewritten in the matrix form as
 
\begin{eqnarray}
\label{potential}
\mathbf{V}=\left(
\begin{array}{ccccc}
 V_{1,1}(z) & V_{1,2}(z) & \ldots & V_{1,N}(z) \\
 V_{2,1}(z) & V_{2,2}(z) & \ldots & V_{2,N}(z) \\
 V_{3,1}(z) & V_{3,2}(z) & \ldots & V_{3,N}(z) \\
 \vdots&\vdots&\ddots&\vdots\\
 V_{N,1}(z) & V_{N,2}(z) & \ldots & V_{N,N}(z) \\
\end{array}
\right).
\end{eqnarray}
To numerically solve the transport problem, we must be able to find $r_{i,j}$, $t_{i,j}$, and $\phi_{i,j}(z)$. By means of the numerical integration of the coupled second-order ordinary differential equations, one needs to know the wave-functions and its derivatives in a certain point of space to be able to propagate the solution in the whole space. The wave-functions described in equations(\ref{chi1}-\ref{chi3}) cannot be univocally determined in a certain point of space because they depend on the $r_{i,j}$, $t_{i,j}$, and $\phi_{i,j}(z)$. Thus, we need to rewrite (\ref{chi3}) in the following way
\begin{eqnarray}
\label{chi3prod}
\chi^{\text{III}}=
\left(
\begin{array}{cccc}
e^{i z k_1}& 0& \ldots & 0\\
0&e^{i z k_2}& \ldots & 0\\
\vdots&\vdots&\ddots&\vdots\\
0&0& \ldots &e^{i z k_N}
\end{array}
\right)\mathbf{t},
\end{eqnarray}
where 
\begin{eqnarray}
\label{transmission_matrix}
\mathbf{t}=
\left(
\begin{array}{ccccc}
t_{1,1}& t_{1,2}& \ldots & t_{1,N}\\
t_{2,1}&t_{2,2}& \ldots & t_{2,N}\\
\vdots&\vdots&\ddots&\vdots\\
t_{N,1}& t_{N,2}& \ldots &t_{N,N}
\end{array}
\right),
\end{eqnarray}
and redefine the wave-functions such as: $\xi^k=\chi^k\mathbf{t}^{-1}$, where $k=$(I,II, or III) and $\mathbf{t}^{-1}$ is the inverse of the transmission matrix given in (\ref{transmission_matrix}). With this redefinition, the wave-function in region III does not depend on the transmission coefficients and it is univocally determined; thereby, enabling us to match the wave-function and its derivative at $z=L/2$, which give the following matrix elements
\[
    \xi^{\text{II}}_{i,j}(L/2)= 
\begin{cases}
    e^{i k_iL/2},& \text{if } i=j\\
    0 ,              & \text{if } i\neq j,
\end{cases}\label{bc1}
\]
\[
   D[\xi^{\text{II}}_{i,j}(z),L/2]= 
\begin{cases}
    i k_ie^{i k_iL/2},& \text{if } i=j\\
    0 ,              & \text{if } i\neq j.
\end{cases}\label{bc2}
\]
Hereafter, to represent the derivative of a function $f(z)$ evaluated at the point $a$, we use the following definition $D[f(z),a]=\left.\frac{df(z)}{dz}\right|_a$ .

Through the matching conditions above, we are able to numerically propagate the solution from $z=L/2$ to $z=-L/2$. Such a propagation can be performed, for example, by means of the 4th order Runge-Kutta method applied to (\ref{hamiltonianchannel}). By using the numerically evaluated $\xi^{\text{II}}(-L/2)$ and $D[\xi^{\text{II}}(z),-L/2]$, we can find $\chi_\text{II}(-L/2)=\xi_\text{II}(-L/2)\mathbf{t}$ and $D[\chi^{\text{II}}(z),-L/2]=D[\xi^{\text{II}}(z),-L/2]\mathbf{t}$. At $z=-L/2$, we also have to impose the boundary conditions $\chi_{\text{II}}(-L/2)=\chi_{\text{I}}(-L/2)$ and $D[\chi^\text{II},-L/2]=D[\chi^\text{I},-L/2]$. These two last equations form a system of linear equations with $N\times N$ equations and variables, which can be solved numerically. If the wave-functions were written in a vectorial form from the beginning, the matching conditions at $z=-L/2$ would lead to a system of linear equations with different number of equations and variables, which is inconsistent. The reflection $r_{i,j}$ and transmission $t_{i,j}$ coefficients are extracted from the solutions to the system of linear equations. The conductance can be evaluated through the Landauer-Buttiker formula $G=G_0\mathrm{Tr}[tt^\dagger]$, where $G_0=e^2/\hbar$ and $t$ is the transmission matrix, whose matrix elements are $t_{i,j}$.

To demonstrate that the numerical scheme works, we have to check if the current density probability is conserved, which can be verified through 

\begin{equation}
 \sum_{i,j=(prop. modes)}\frac{v_i}{v_j}|r_{i,j}|^2+\frac{v_i}{v_j}|t_{i,j}|^2=1,
 \end{equation} 
 where $v_i=\hbar k_i/m^*$ and the sum above should be evaluated over the propagating modes, \textit{i.e.}, only if $k_i$ is real. The evanescent modes, characterized by the imaginary $k_i$, have an important role in the transport properties but do not contribute to the conservation of the current density probability. 
 \begin{figure}[t]
    \includegraphics[width=0.45\textwidth]{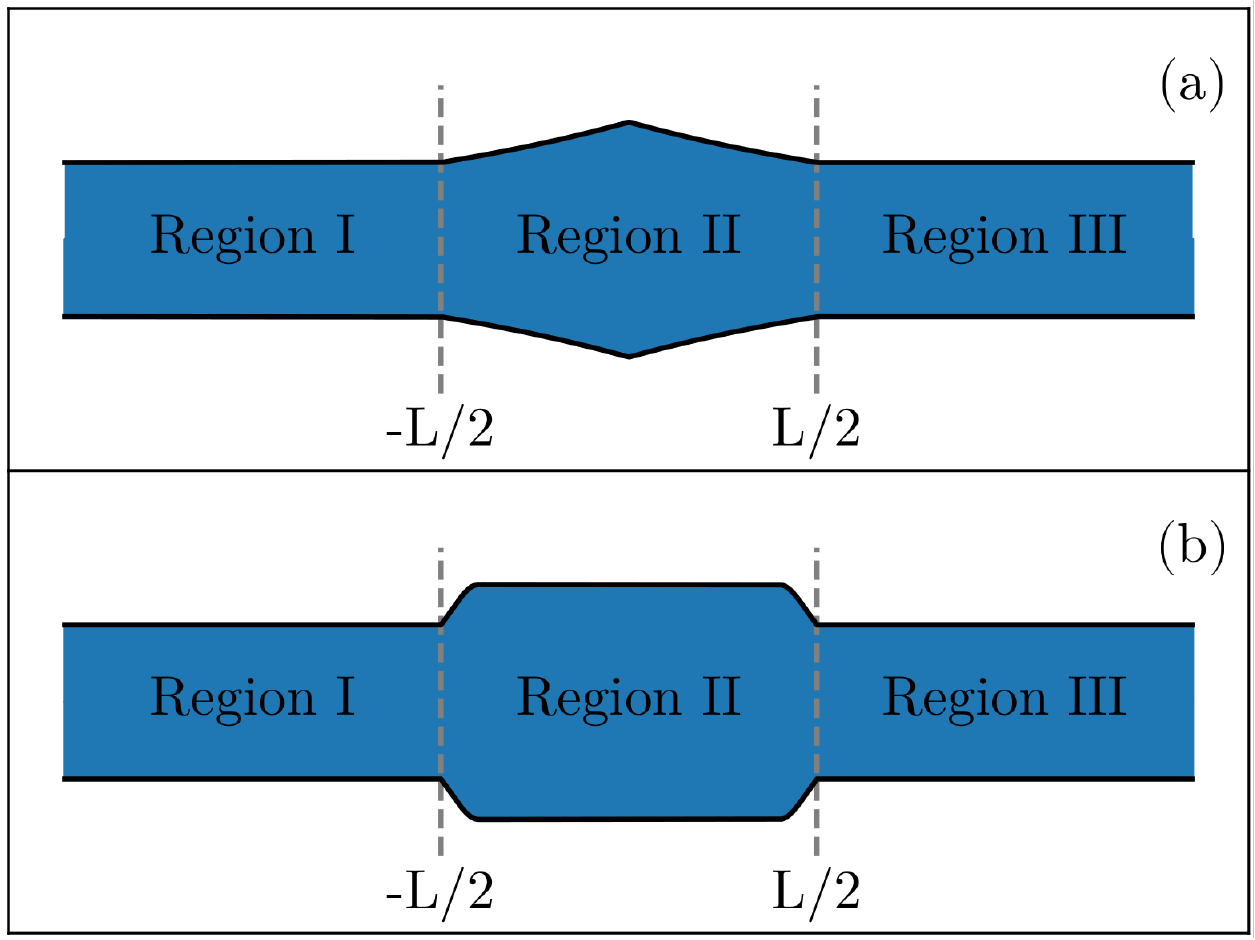}
    \caption{Schematic picture showing the longitudinal cut of the NW containing an expansion of the radius in the region II with length $L$. Panel (a) shows the cone-shape deformation $\Delta r_1(z)$, while panel (b) shows the square-shape deformation $\Delta r_2(z)$. }
    \label{fig:nanowire}
\end{figure}
 By employing this numerical scheme, we can investigate the NW containing a variable radius within the region II. We model the z-dependence of the radius by the function $r(z)=r_0+\Delta r(z)$, where $r_0$ is the NW radius and $\Delta r(z)$ is the variation of the radius within region II.
We choose two different forms for the radius variation, $\Delta r_1(z)=\Delta r_0(1-2|z|/L)$ for $|z|\leq L/2$ and 
\begin{eqnarray}
\Delta r_2(z)=\Delta r_0
\begin{cases}
    \sin{\left(\frac{\pi (z+L/2)}{2L_0}\right)},\;-\frac{L}{2}\leq z< -\frac{L}{2}+L_0\\
    1, \qquad\qquad\quad-\frac{L}{2}+L_0\leq z\leq \frac{L}{2}-L_0\\
    \sin\left(\frac{\pi (L/2-z)}{2L_0}\right),\;\frac{L}{2}-L_0< z\leq \frac{L}{2}
\end{cases}\nonumber
\end{eqnarray}
where $\Delta r_0$ is the maximum variation of the radius at $z=0$ and $L_0$ defines the the width of the ramp between zero and $\Delta r_0$. The first form $\Delta r_1(z)$ describes a NW with a cone-shape deformation (panel (a) of figure 1), while the second form $\Delta r_2(z)$ produces a square-shape deformation (panel (b) of figure 1). We choose these two forms to probe the effects of different shapes of deformations in the electronic transport properties. The main difference between both shapes is that the square-shape deformation has a bigger area than the cone-shape deformation.

To find analytical solutions for the transverse modes, we employ the harmonic oscillator potential to describe the lateral confining potential $V_c(x,y)=m^*\omega_0^2(x^2+y^2)/2$, where $\omega_0=\hbar/(m^*r_0^2)$. These transverse modes are solutions of the following equation: 
\begin{equation}
\left[\frac{-\hbar^2\nabla^2_\perp}{2m^*}+\frac{m^*\omega_0^2}{2}(x^2+y^2)-E_{n,m}\right]\Phi_{n,m}(x,y)=0.\label{hamiltoniant_ho}
\end{equation}
 By using creation and annihilation operators:
\begin{equation}
a_q=\sqrt{\frac{m^*\omega_0}{2\hbar}}q+\frac{i}{\sqrt{2m^*\hbar\omega_0}}p_q,
\end{equation}
\begin{equation}
a^\dagger_q=\sqrt{\frac{m^*\omega_0}{2\hbar}}q-\frac{i}{\sqrt{2m^*\hbar\omega_0}}p_q,
\end{equation}
where $q=x,y$ and $p_q=-i\hbar\frac{\partial}{\partial q}$, we can rewrite (\ref{hamiltoniant_ho}) as
\begin{equation}
\hbar\omega_0\left[a_x^\dagger a_x+a_y^\dagger a_y+1\right]|\Phi_{n,m}\rangle=E_{n,m}|\Phi_{n,m}\rangle,\label{hamiltoniant_ho1}
\end{equation}
whose eigenvalues are $E_{n,m}=\hbar\omega_0(n+m+1)$, where $n=m=0,1,2,\ldots$, where $n$ ($m$) are associated to the number operator $N_x=a_x^\dagger a_x$ ($N_y=a_y^\dagger a_y$). In region II, we have a change in the radius that causes a change in the confinement, which can be described by $\omega(z)=\frac{\hbar}{m^*r(z)^2}$.
 Therefore, the scattering potential can be modelled as $V(x,y,z)=m^*[\omega(z)^2-\omega_0^2](x^2+y^2)/2$ , whose matrix elements (\ref{potential}) are given by
\begin{eqnarray}
&&V_{n^\prime,m^\prime}^{n,m}(z)=v(z)\langle\Phi_{n,m}|(a_x+a_x^\dagger)^2+(a_y+a_y^\dagger)^2|\Phi_{n^\prime,m^\prime}\rangle\nonumber\\&&=v(z)\left[d^n_{m}\delta_{n,n^\prime}\delta_{m,m^\prime}+c^n_{n^\prime}\delta_{m,m^\prime}+c^m_{m^\prime}\delta_{n,n^\prime}\right]\label{vmatrixelements},
\end{eqnarray}
where $v(z)=\frac{\hbar[\omega(z)^2-\omega_0^2]}{4\omega_0}$, $d^n_{m}=(2n+2m+1)$, and $c^n_{n^\prime}=\left(\sqrt{n^\prime(n^\prime-1)}\delta_{n,n^\prime-2}+\sqrt{(n^\prime+1)(n^\prime+2)}\delta_{n,n^\prime+2}\right)$.
In (\ref{vmatrixelements}), there are the diagonal terms proportional to $V^d_{n,m}(z)=d^n_mv(z)$ and the off-diagonal terms that couples channels that obey the selection rules $n=n^\prime\pm 2$ with $m=m^\prime$ or $m=m^\prime\pm 2$ with $n=n^\prime$. The diagonal terms can be rewritten as a function of the radius variation,  $V^d_{n,m}(z)=d^n_m\hbar\omega_0\frac{[r_0^4-r(z)^4]}{4r(z)^4}$; thereby, showing that it is an attractive potential if the radius within the region II increases ($r(z)\geq r_0$). For a narrowed NW in the region II, we have that $r(z)\leq r_0$ and the diagonal term becomes a repulsive potential.

\section{Numerical Results}

\subsection{Deformation effects}
We use the following channels in all numerical calculations using a variation in the quantum number $m$: $\{ (0,0,\pm), (0,1,\pm), (0,2,\pm), (0,3,\pm) \}$, which is a good approximation for Fermi energy $E\leq 4\varepsilon_0$, where $\varepsilon_0=\hbar\omega_0$. The channels with variation in the quantum number $n$, such as $\{(0,0,\pm), (1,0,\pm), (2,0,\pm), (3,0,\pm)\}$, are decoupled from the channels where the variation occurs only in $m$; therefore, the conductance considering the change of both quantum numbers $(n,m)$ is simply obtained by multiplying the results from the variation in the quantum number $m$ by two. We start our analysis by exploring the effect of a narrowed NW ($\Delta r_0<0$). In figure~(2), we plot the normalized conductance $G/G_0$ as a function of the normalized Fermi energy $E/\varepsilon_0$ for different values of the $\Delta r_0$ and for a fixed size of the region II ($L=8r_0$), considering both shapes of deformation. 
By increasing the value of $|\Delta r_0|$, one can notice that (i) the Fermi energy where the conductance becomes different from zero shifts towards higher energies as the compression is increased and (ii) the conductance does not exhibit dips and it is smoothed for the cone-shape deformation. Both results are expected because the appearance of the repulsive potential in the diagonal terms of (\ref{vmatrixelements}) for a narrowed NW, which becomes stronger for higher values of $|\Delta r_0|$. The conductance for square-shape deformation (dashed lines in figure~(2)) resembles the results for a square barrier potential, as would be expected.

\begin{figure}[th!]
    \includegraphics[width=0.45\textwidth]{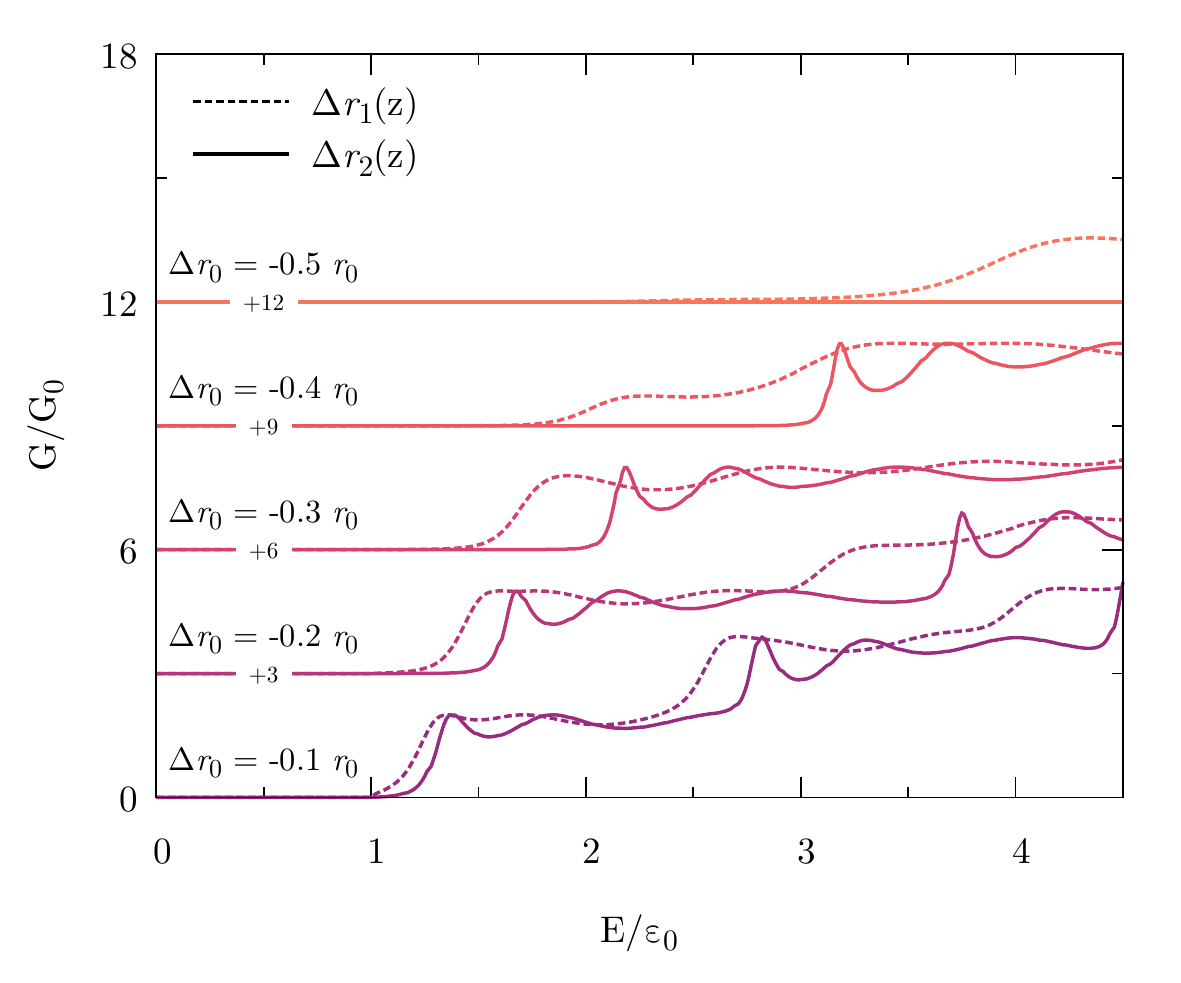}
    \caption{Normalized conductance as a function of the normalized Fermi energy for different radius compression $\Delta r_0/r_0=-0.1,-0.2,-0.3,-0.4,$ and -0.5, considering the length $L/r_0 = 8.0$. Solid (dashed) lines describe the results for the square (cone)-shape deformation model. The curves are offset for clarity according to the indicated values.}
    \label{fig:expansion}
\end{figure}

\begin{figure}[bh!]
    \includegraphics[width=0.45\textwidth]{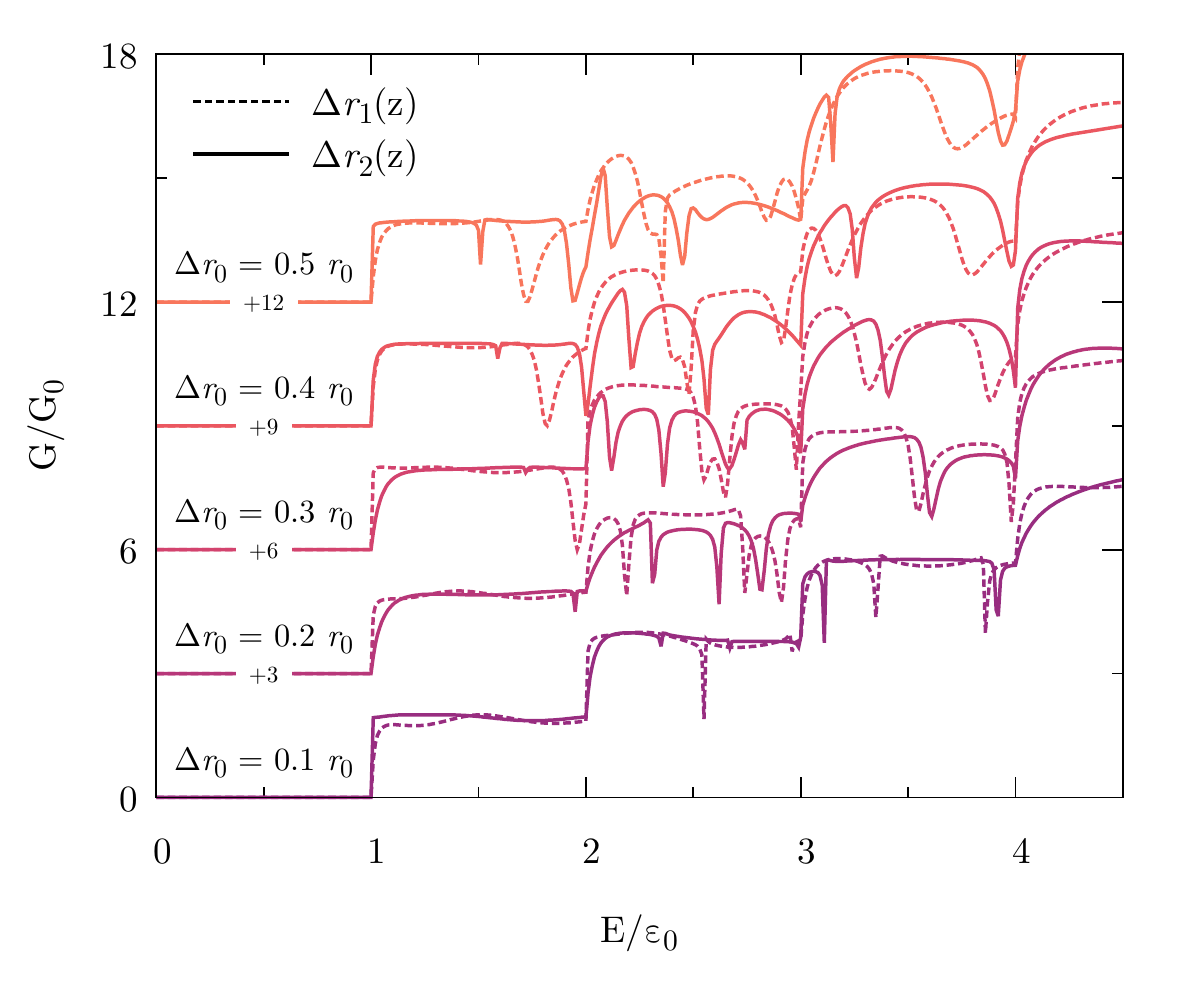}
    \caption{Normalized conductance as a function of the normalized Fermi energy for different radius expansion $\Delta r_0/r_0=0.1,0.2,0.3,0.4,$ and 0.5, considering the length $L/r_0 = 8.0$. Solid (dashed) lines describe the results for the square (cone)-shape deformation model. The curves are offset for clarity according to the indicated values.}
    \label{fig:contraption}
\end{figure}

On the other hand, dips can be observed when an expansion occurs in the radius of the NW in region II. In figure~(3), the normalized conductance is plotted as a function of the normalized Fermi energy considering $\Delta r_0>0$ for $L=8r_0$. The conductance is almost constant within the first plateau ($\varepsilon_0 < E < 2 \varepsilon_0$) and the dip is only observed for $\Delta r_0 \ge 0.3r_0$ for the cone-shape deformation (dashed curves in figure~(3)). However, many dips are present for an expanded NW for $E > 2 \varepsilon_0$, which indicates strong reflection of electrons, manifesting the reentrance behaviour within the second and third plateaus. As the value of $\Delta r_0$ increases the position of the reentrance changes in the normalized energy axis as can be observed in figure~(3). The square-shape deformation presents more dips in the conductance than the cone-shape deformation because the dips can be related to quasi-bound states localized in region II~\cite{Gurvitz} that interfere to the scattering wave-functions; thus, the wider the potential, the higher the number of quasi-bound states. To understand the role of the size of the region II, we fix $\Delta r_0=0.3 r_0$ and we vary the length $L=4.4, 5.6, 6.8, 8.0,$ and$, 9.2 r_0$. Such results are shown in figure~(4). For $L=$ 4.4 and 5.6$r_0$, the dip in the first plateau is absent while two dips are present within the second plateau, for the cone-shape deformation (dashed curves in figure~(4)). On the other hand, the dip in the first plateau appears for $L=$4.4 and 5.6$r_0$ when the square-shape deformation is considered (solid curves in figure~(4)). For $L=$6.8, 8.0, and 9.2$r_0$, a dip in the first plateau is present for $\Delta r_1(z)$ (dashed curves in figure(4)). On the contrary, the dip in the first plateau becomes very narrow for $L=$6.8, 8.0, and 9.2$r_0$ when $\Delta r_2(z)$ is employed (solid curves in figure~(4)).
It can be noted that the number of dips within second and third plateaus are really influenced by the length $L$ due to interference phenomena occurring between quasi-bound and scattering sates~\cite{Gurvitz}. The results of this section demonstrate that the reentrance feature can indeed be observed even in the absence of the SOC and the role of deformations in the transport properties in NWs.

\begin{figure}[th!]
    \includegraphics[width=0.45\textwidth]{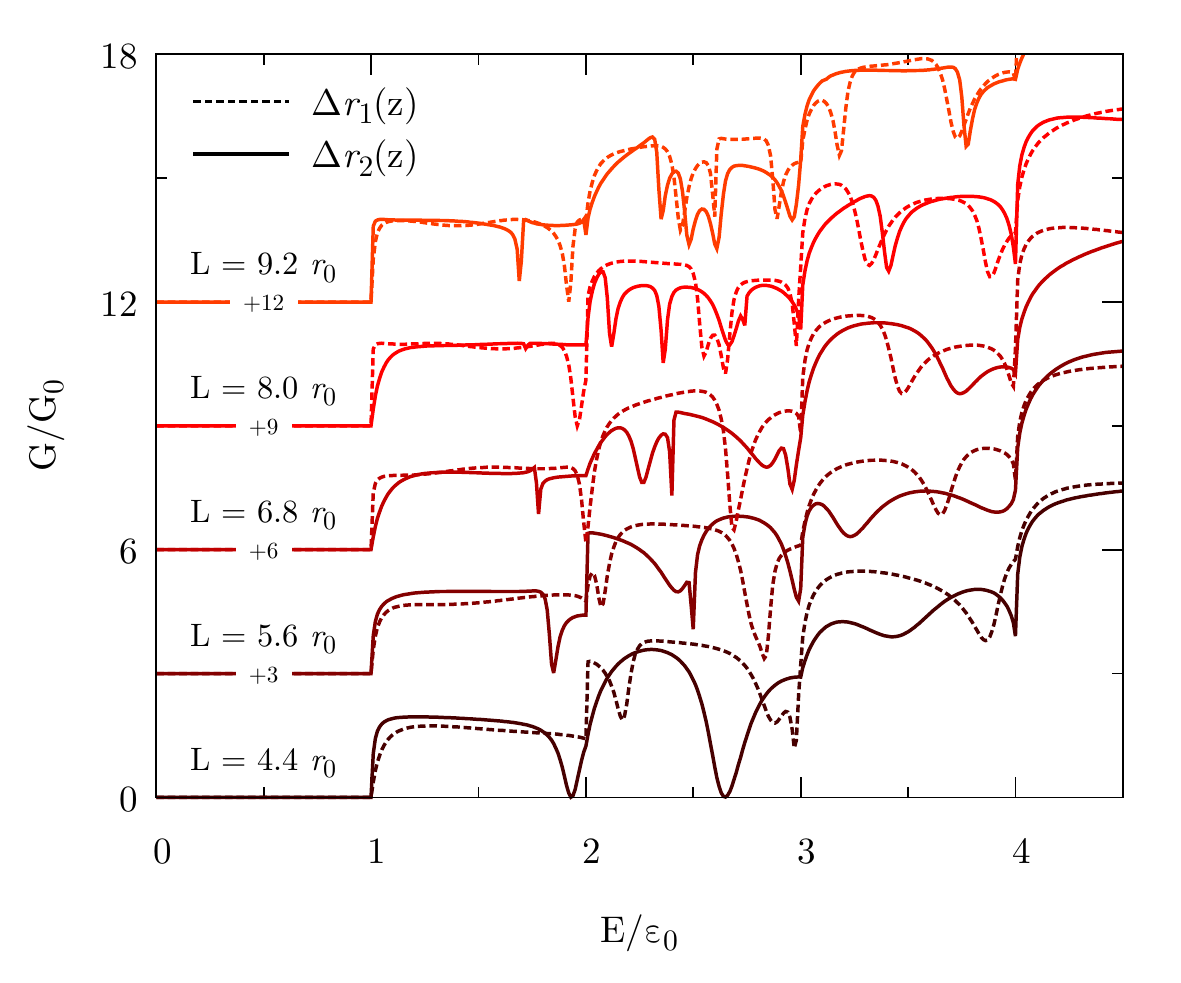}
    \caption{Normalized conductance as a function of the normalized Fermi energy for different length $L/r_0=4.4,5.6,6.8,8.0,$ and 9.2, and fixed radius deformation $\Delta r_0 / r_0 = 0.3$. Solid (dashed) lines describe the results for the square (cone)-shape deformation model. The curves are offset for clarity according to the indicated values.}
    \label{fig:legnth}
\end{figure}

\subsection{Rashba SOC}
In this section, we want to explore the role of the structural deformation in the NW when the Rashba SOC is also taken into account. In this case, we must add the Rashba term to (1), which is given by the following equation
\begin{equation}
    H_{SO} = \alpha(z) (k_y \sigma_z - k_z \sigma_y),
\end{equation}
where $\sigma_y$ ($\sigma_z$) and $k_y$ ($k_z$) are respectively the Pauli spin matrix and the wave vector in the y-direction (z-direction). 
 We assume that the Rashba interaction only acts in the region II of length L along the z-direction; thus, $\alpha(z)=\alpha\left[\Theta(z+L)-\Theta(z-L)\right]$ and $\Theta(z)$ is the Heaviside function and $\alpha$ is the Rashba coupling strength. The reason for considering a localized Rashba SOC is related to experimental setups where an electric field is applied between top and bottom gates, which induces the  emergence of a structural inversion asymmetry SOC~\cite{Moroz}.
 We must explicitly add the spin degree of freedom when the Rashba SOC is considered; therefore, the matrix elements evaluated in (\ref{vmatrixelements}) must be multiplied by $\delta_{s,s'}$, where $s,s'\in \{-,+\}$. Using the transverse modes, which are solutions of (2), we can calculate
the matrix elements for the Rashba SOC, as follows
\begin{equation}
    H_{j,j'}^{SO} = -is\left[ b^m_{m^\prime}  \delta_{n, n'} \delta_{s,s'} +\{ \alpha(z),k_z \} \delta_{n, n'} \delta_{m, m'}\delta_{s,-s'} \right],
\end{equation}
with $b^m_{m^\prime}=\frac{\sqrt{2}\alpha(z)}{r_0}\left[ \sqrt{m'}\delta_{m,m'-1} - \sqrt{m'+1}\delta_{m, m'+1} \right]$, $\{\alpha(z),k_z\} = (\alpha(z)k_z + k_z\alpha(z))/2$, and $j=(n,m,s)$.

\begin{figure}[th!]
    \includegraphics[width=0.45\textwidth]{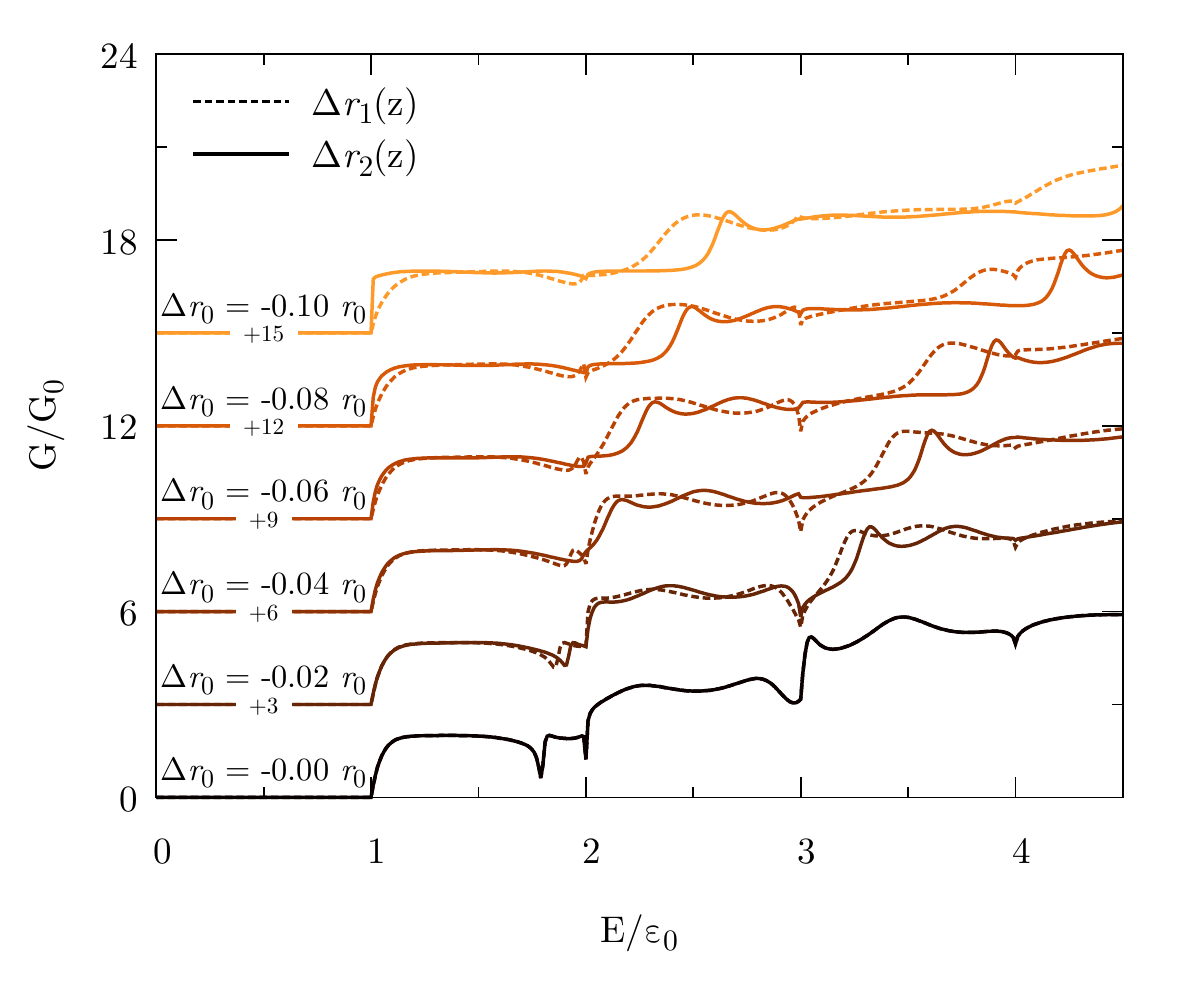}
    \caption{Normalized conductance as a function of the normalized Fermi energy for different radius compression $\Delta r_0/r_0=0,-0.02,-0.04,-0.06,-0.08,$ and -0.1, considering the length $L/r_0 = 8.0$ and the Rashba constant $\alpha/\alpha_0 = 0.4$. Solid (dashed) lines describe the results for the square (cone)-shape deformation model. The curves are offset for clarity according to the indicated values.}
    \label{fig:rashba_constriction}
\end{figure}

\begin{figure}[t!]
    \includegraphics[width=0.45\textwidth]{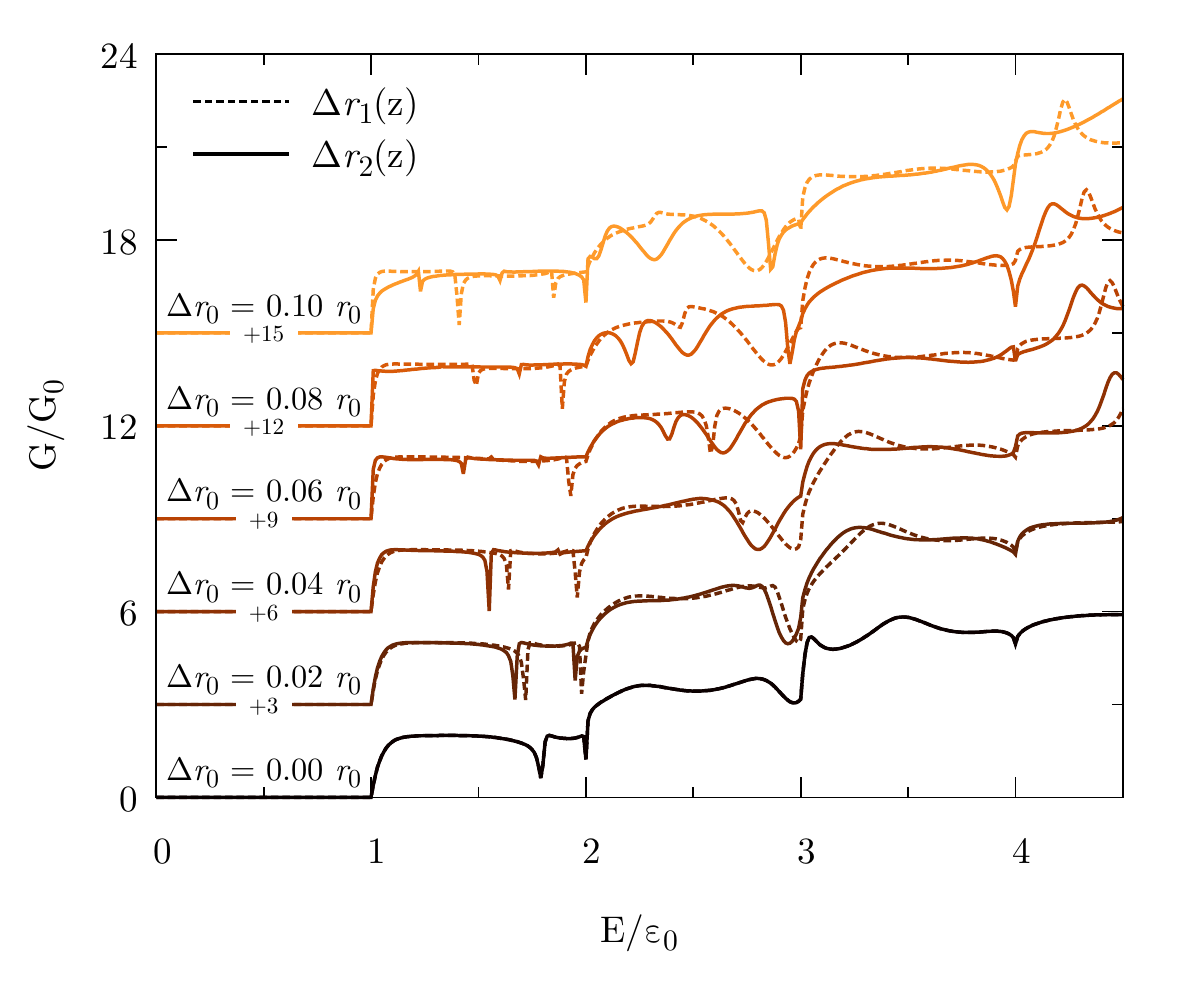}
    \caption{Normalized conductance as a function of the normalized Fermi energy for different radius expansion $\Delta r_0/r_0=0,0.02,0.04,0.06,0.08$ and 0.1, considering the length $L/r_0 = 8.0$  and the Rashba constant $\alpha/\alpha_0 = 0.4$. Solid (dashed) lines describe the results for the square (cone)-shape deformation model. The curves are offset for clarity according to the indicated values.}
    \label{fig:rashba_expansion}
\end{figure}

\begin{figure}[th!]
    \includegraphics[width=0.45\textwidth]{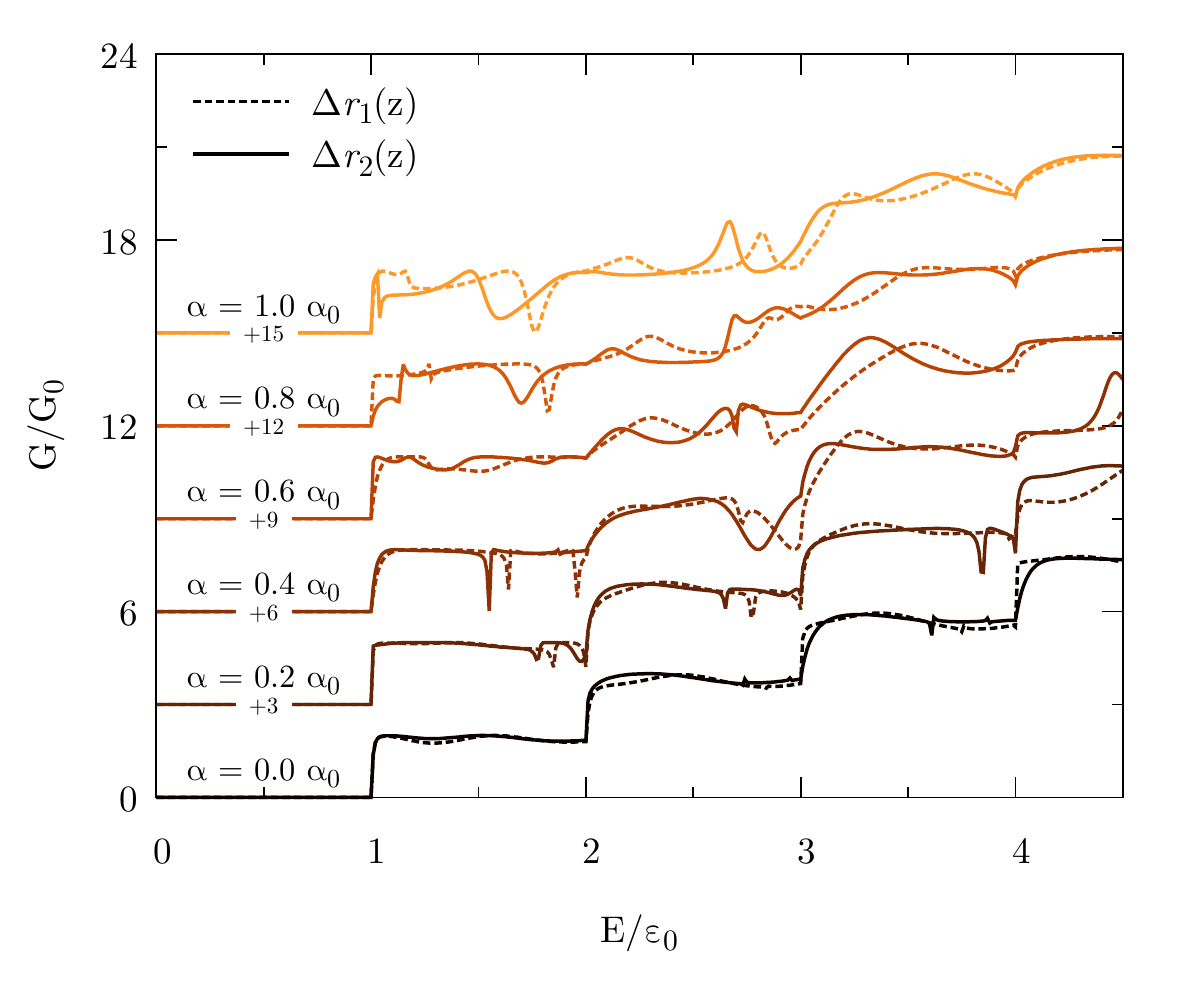}
    \caption{Normalized conductance as a function of the normalized Fermi energy for a fixed value of the radius expansion $\Delta r_0 / r_0=0.04$ and length $L/r_0 = 8.0$, considering the variation of the Rashba constant $\alpha/\alpha_0=0.0, 0.2, 0.4, 0.6, 0.8$, and 1.0. The curves are offset for clarity according to the indicated values. Solid (dashed) lines describe the results for the square (cone)-shape deformation model.}
    \label{fig:rashba_alpha}
\end{figure}

Due to the Rashba SOC, the boundary conditions for the derivative of wave-functions must be reformulated as~\cite{Xiao}: $D[\chi^\text{II},-L/2]=D[\chi^\text{I},-L/2] +\frac{i\mathbf{M}}{2}\chi_{II}(L/2)$ and $D[\chi^\text{II},L/2]=D[\chi^\text{I},L/2] -\frac{i\mathbf{M}}{2}\chi_{II}(L/2)$, where $\mathbf{M}$ is a matrix, whose elements are given by $\mathbf{M}_{i,j}=-2 \alpha \delta_{n,n'} \delta_{m, m'} \delta_{s,s'}$.
Results including the SOC are shown in Figs.~(\ref{fig:rashba_constriction}-\ref{fig:rashba_alpha}). In figure~(\ref{fig:rashba_constriction}), we plot the normalized conductance as a function of the Fermi energy considering $\alpha=0.4\alpha_0$ ($\alpha_0=\varepsilon_0r_0$) and for different values for the radius compression, $\Delta r_0/r_0 = 0.0,-0.02,-0.04,\ldots,-0.1$. When $\Delta r_0 = 0.0$, there are two dips in the first plateau at $E\approx1.8\varepsilon_0$ and $E\approx2.0\varepsilon_0$ and one dip in the second plateau at $E\approx3.0\varepsilon_0$ only due to the Rashba SOC. For $\Delta r_0 = -0.02r_0$, all dips in the conductance are reduced. By further compressing the radius of the NW, one can notice in figure~(\ref{fig:rashba_constriction}) that the dips completely disappear for $\Delta r_0 \ge -0.04 r_0$ when the square-shape model is considered (solid curves in figure 5). These results are expected because the local radius compression works as a repulsive potential, which can cancel the attractive potential given by the SOC~\cite{Sanchez}. When the radius of the NW suffers a local expansion in the same region where the SOC takes place, we have the appearance of more dips and a shift towards smaller energies as the expansion increases, as can be seen in figure~(\ref{fig:rashba_expansion}). In figure~(\ref{fig:rashba_alpha}), we plot the normalized conductance as a function of the Fermi energy considering $\Delta r_0 = 0.04 r_0$ and different values for the Rashba constant ($\alpha/\alpha_0=0.0-1.0$ with step 0.2). For these parameters, there is no dip in the first plateau for $\alpha=0.0$, but dips appear for $\alpha\geq0.2\alpha_0$. One can also notice that both deformation models present different results, even considering a small deformation $\Delta r_0 = 0.04 r_0$. All results presented in  Figs.~(\ref{fig:rashba_constriction}-\ref{fig:rashba_alpha}) demonstrate that there is a competition between Rashba SOC and structural deformations, which can cause a misinterpretation of experimental data when related to the reentrant feature. 
\subsection{Magnetic Field}
\begin{figure}[th!]
 \includegraphics[width= 0.45\textwidth]{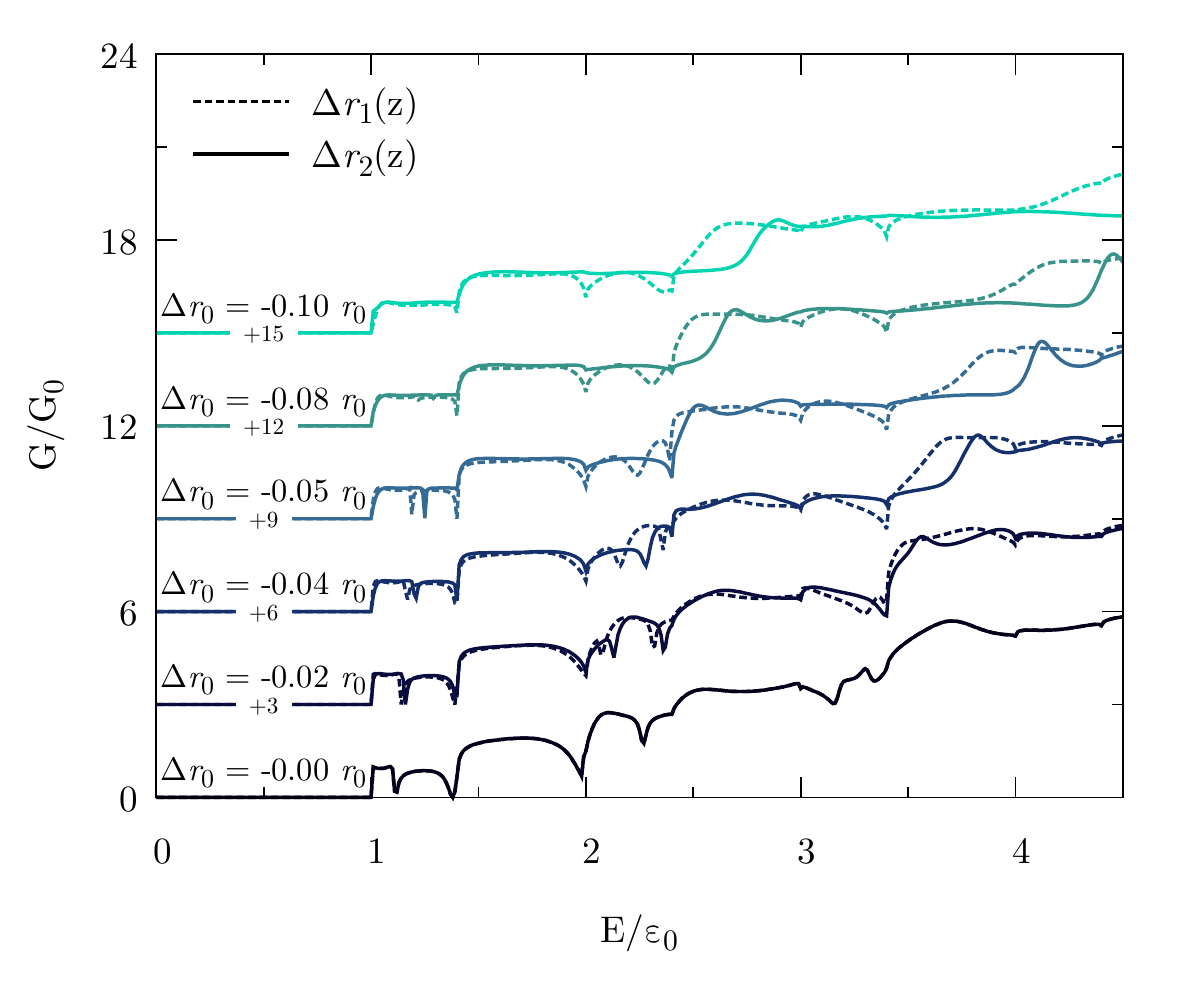}
    \caption{Normalized conductance as a function of the normalized Fermi energy for different radius compression $\Delta r_0/r_0=0.0,-0.02,-0.04,-0.06,-0.08,$ and -0.1, considering the length $L/r_0 = 8.0$, the Rashba constant $\alpha/\alpha_0=0.4$,  and Zeeman energy $E_Z / \varepsilon_0 =0.2$. The curves are offset for clarity according to the indicated values. Solid (dashed) lines describe the results for the square (cone)-shape deformation model.}
    \label{fig:rashba_B_constriction}
\end{figure}

\begin{figure}[t!]
    \includegraphics[width=0.45\textwidth]{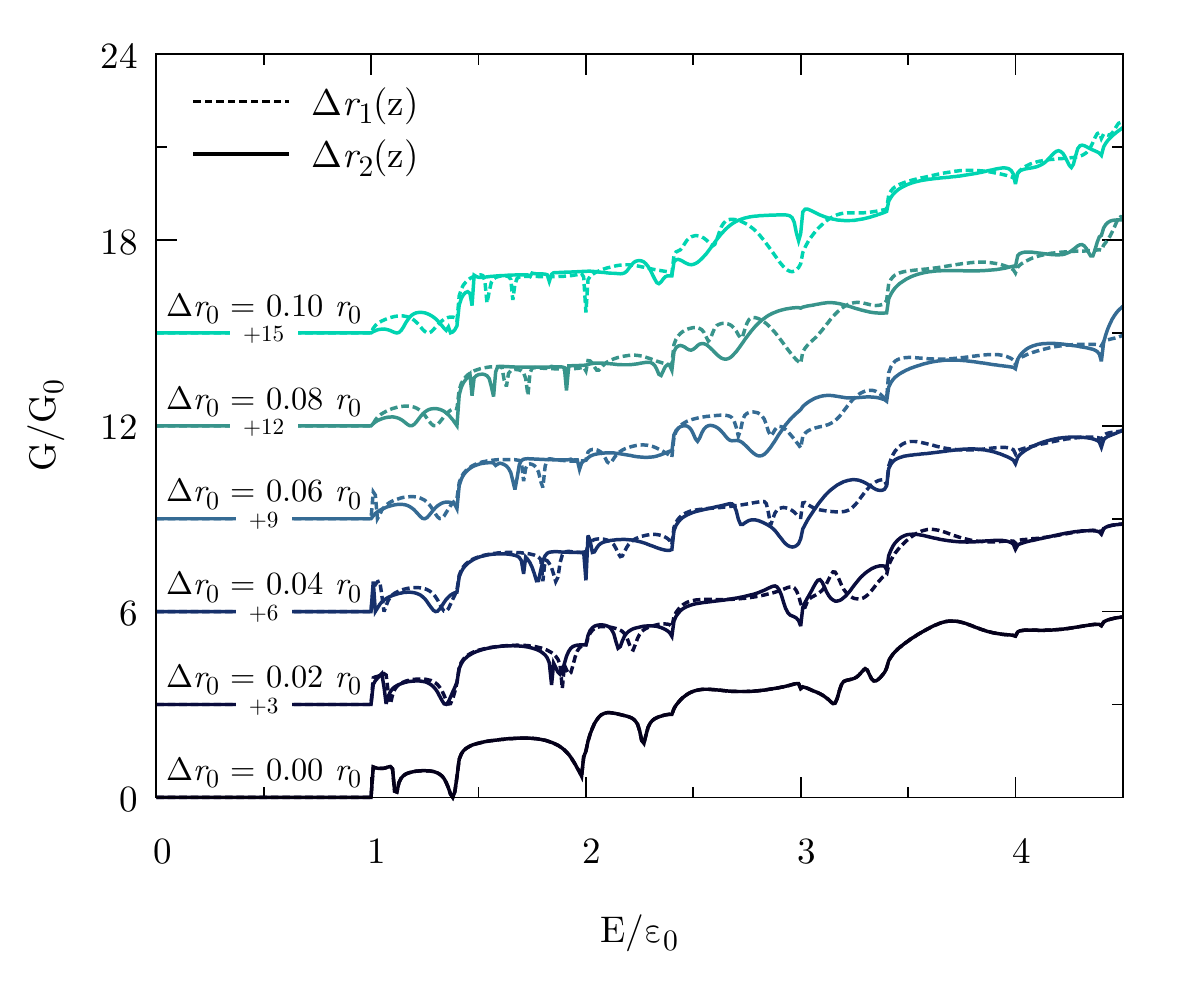}
    \caption{Normalized conductance as a function of the normalized Fermi energy for different radius expansion $\Delta r_0/r_0=0.0,0.02,0.04,0.06,0.08,$ and 0.1, considering the length $L/r_0 = 8.0$, the Rashba constant $\alpha/\alpha_0=0.4$, and Zeeman energy $E_Z / \varepsilon_0 = 0.2$. The curves are offset for clarity according to the indicated values. Solid (dashed) lines describe the results for the square (cone)-shape deformation model. }
    \label{fig:rashba_B_expansion}
\end{figure}
\begin{figure}[ht!]
    % lenght variation linear
    \includegraphics[width=0.45\textwidth]{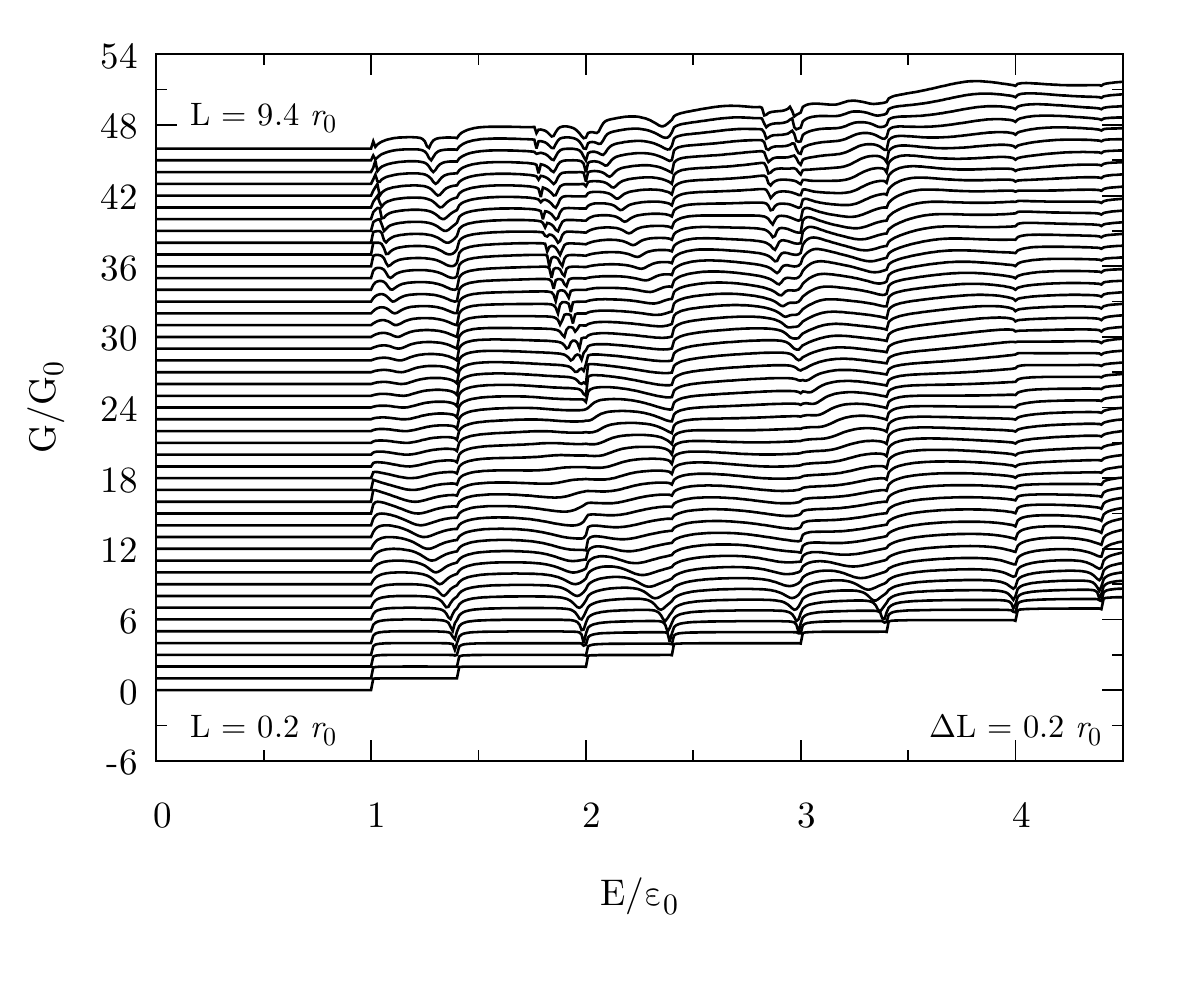}
    \caption{Normalized conductance as a function of the normalized Fermi energy for different length $L/r_0 = 0.2$ to $9.4$ with step $\Delta L/r_0 = 0.2$, considering $\Delta r_1(z)$ for the radius variation function with $\Delta r_0/r_0 = 0.04$, the Rashba constant $\alpha/\alpha_0 = 0.4$, and Zeeman energy $E_Z/ \varepsilon_0 = 0.2$.  The curves are offset for clarity.}
    \label{fig10}
\end{figure}

\begin{figure}[ht!]
    % rashba variation linear
    \includegraphics[width=0.45\textwidth]{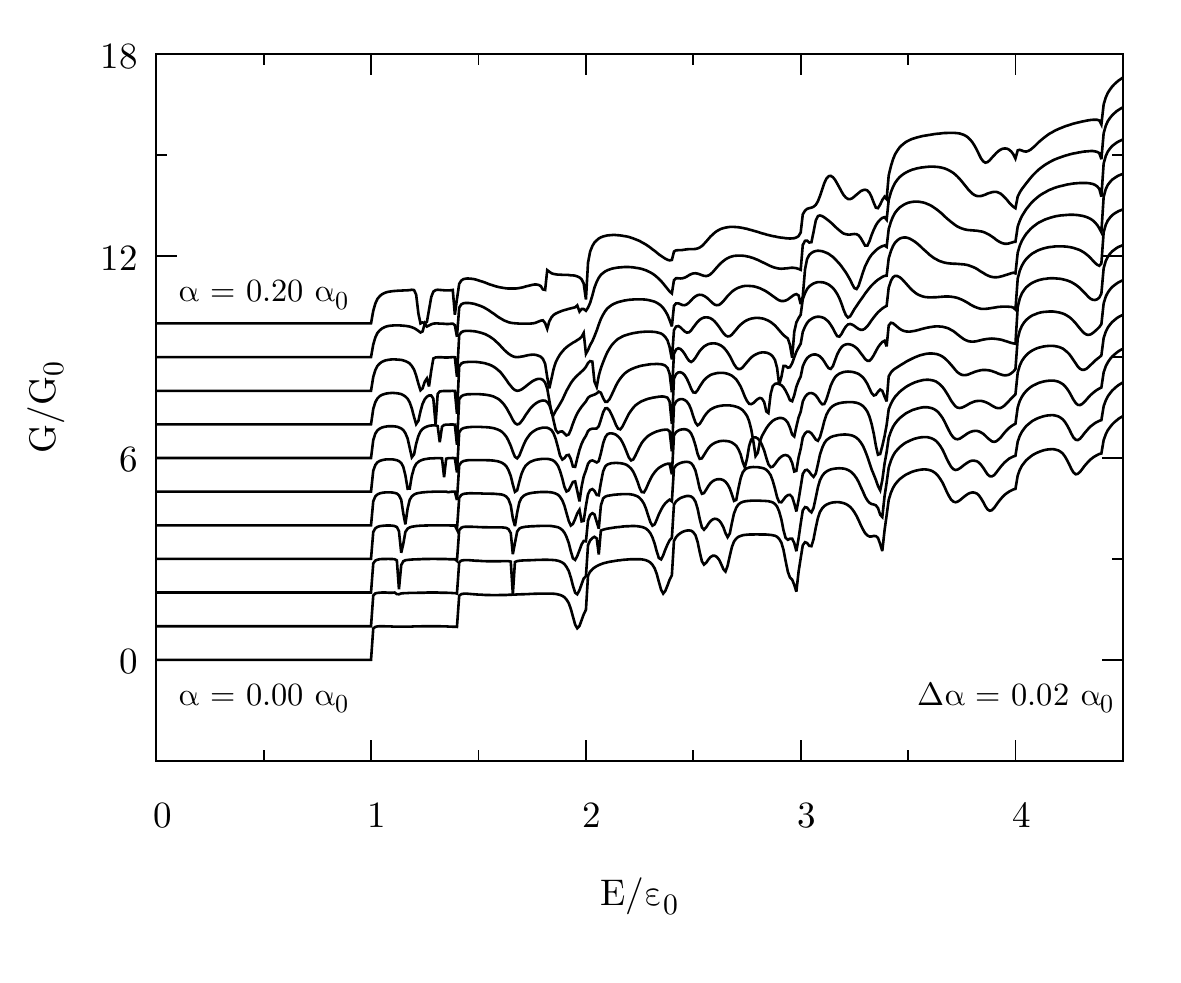}
    \caption{Normalized conductance as a function of the normalized Fermi energy for different Rashba constant $\alpha / \alpha_0 = 0.0$ to $0.2$ with step $\Delta \alpha / \alpha_0 = 0.02$, considering $\Delta r_1(z)$ for the radius variation function with $\Delta r_0/r_0 = 0.3$, length $L/r_0 = 8.0$, and Zeeman energy $E_Z / \varepsilon_0 = 0.2$.  The curves are offset for clarity.}
    \label{fig11}
\end{figure}

\begin{figure}[ht!]
    % Field variation plateau (alpha = 0.20)
    \includegraphics[width=0.45\textwidth]{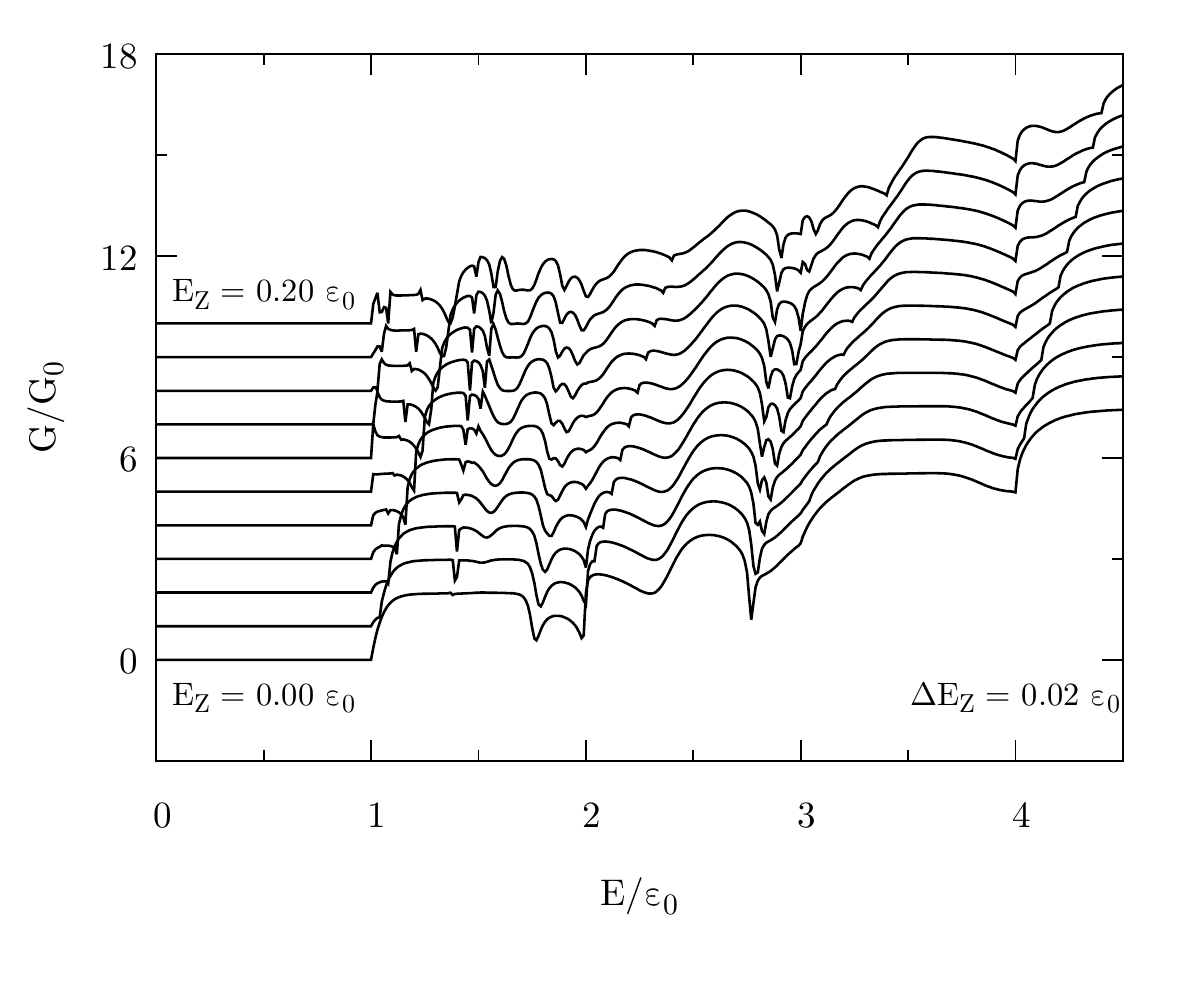}
    \caption{Normalized conductance as a function of the normalized Fermi energy for different Zeeman energy $E_Z / \varepsilon_0 = 0.0$ to $0.2$ with step $\Delta E_Z / \varepsilon_0 = 0.02$, considering $\Delta r_2(z)$ for the radius variation function with $\Delta r_0/r_0 = 0.3$, the Rashba constant $\alpha/\alpha_0 = 0.2$, and length $L/r_0 = 8.0$. The curves are offset for clarity.}
    \label{fig12}
\end{figure}

\begin{figure}[t!]
    % Field variation plateau (alpha = 0.80)
    \includegraphics[width=0.45\textwidth]{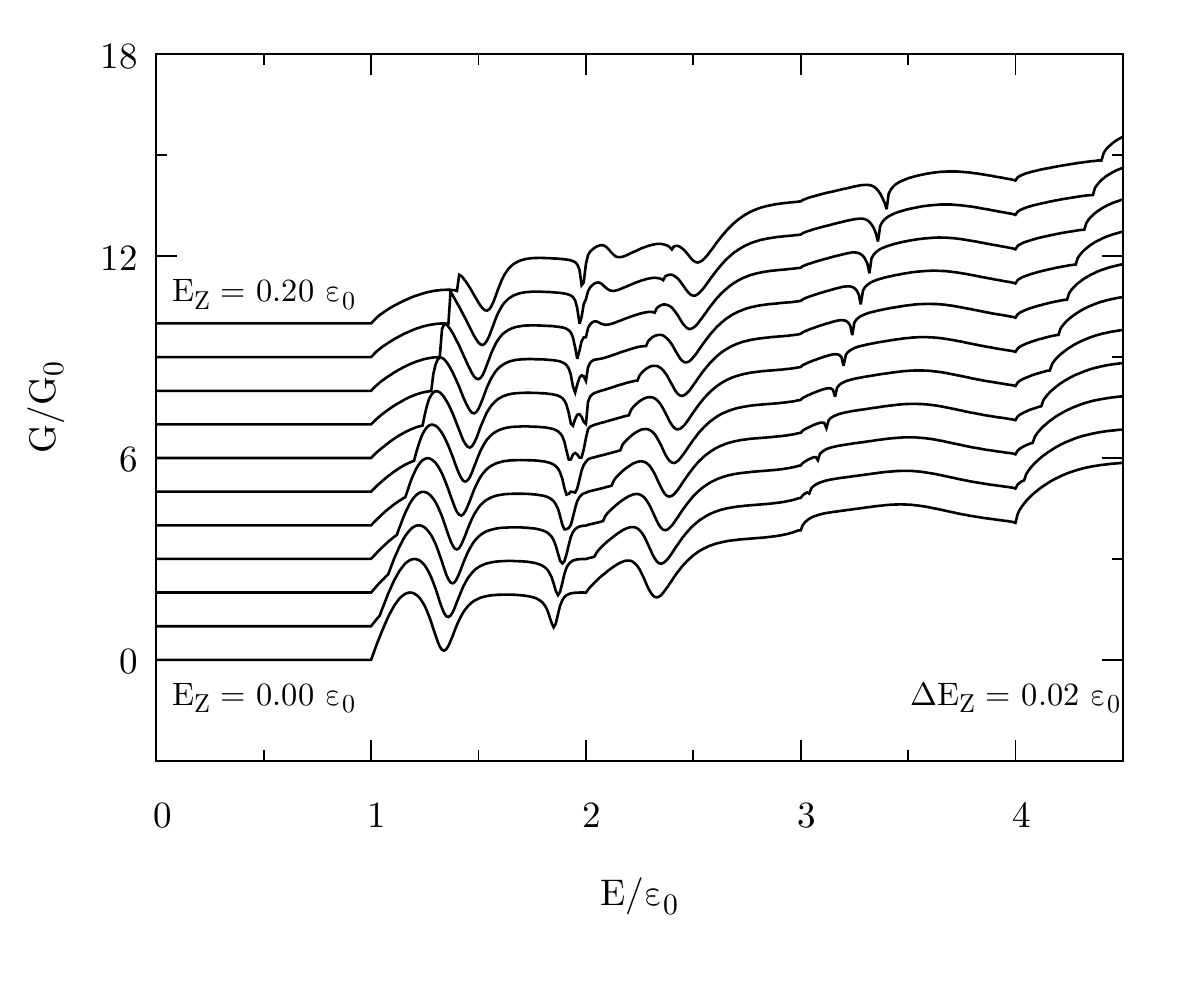}
    \caption{Normalized conductance as a function of the normalized Fermi energy for different Zeeman energy $E_Z / \varepsilon_0 = 0.0$ to $0.2$ with step $\Delta E_Z / \varepsilon_0 = 0.02$, considering $\Delta r_2(z)$ for the radius variation function with $\Delta r_0/r_0 = 0.3$, the Rashba constant $\alpha/\alpha_0 = 0.8$, and length $L/r_0 = 8.0$. The curves are offset for clarity.}
    \label{fig13}
\end{figure}

Ideal helical states occur when an external magnetic field is applied to the NW. In this section, we add a uniform magnetic field applied in the x-direction to the previous NW configuration, in accordance to the experiment of \onlinecite{Heedt}. In this case, we must add the Zeeman Hamiltonian $H_Z=E_Z\sigma_x$ to (\ref{hamiltonian}), where $E_Z$ is the Zeeman energy. The matrix elements for this Hamiltonian are given by $H ^Z_{j,j'}=E_Z\delta_{n,n'}\delta_{m,m'}\delta_{s,-s'}$. When the magnetic field is included, the subband energy split as $E^\pm_{n,m}=E_{n,m}\pm E_Z$. This would move the energy position where the first step in the conductance takes place. Yet, the electron source will be affected by the uniform magnetic field in the same way, thereby canceling out this energy shift. Thus, we have compensated for this difference and considered that the total energy where the first channel opens does not change with the variation of the magnetic field. In figure~8, we plot the normalized conductance for a fixed Zeeman energy, $E_Z=0.2\varepsilon_0$, as a function of the total energy considering $\alpha=0.4\alpha_0$, $L=8r_0$ and $\Delta r_0/r_0=$0.0,-0.02,-0.04,-0.06,-0.08,and -0.1. In this case, many dips appear when $\Delta r_0=$0.0, but they fade out even for a small compression $\Delta r_0=$-0.04 $r_0$. By further compressing the NW, we clearly observe the disappearance of the dips for $\Delta r_0=$-0.08 $r_0$ using the square-model (solid curves in figure~8). On the other hand, more dips appear for a expanded NW due to interference phenomenon between scattering and quasi-bound states~\cite{Gurvitz}, as can be observed in figure 9. Such results demonstrate how a small structural deformation can play an important role in the observation of the reentrant feature in NWs when a magnetic field and Rashba SOC coexist. To further explore our method, we consider some extra cases to understand the dependence of the dips in the conductance as a function of different parameters. First, we vary the length of the region II within the range $L\in[0.2,9.4]r_0$ with step $\Delta L=0.2r_0$, assuming $\alpha/\alpha_0=0.4$, $E_Z=0.2\varepsilon_0$, and the cone-shape deformation with $\Delta r_0/r_0=0.04$, whose results for the conductance are shown in figure~10. As already discussed, the Rashba SOC works as an attractive potential~\cite{Sanchez}, specifically as a quantum well with depth $V_0=-m\alpha^2/\hbar^2$ and length $L$. When $L\leq0.6r_0$ and $\alpha=0.4\alpha_0$ there is no dip in the conductance because there is no quasibound state within this quantum well. According to the approximate two-channel model described in Ref.~\onlinecite{Gurvitz}, the dip in the conductance occurs when an attractive potential induces the formation of a quasi-bound state. In figure 10 there is only one dip in the plateaus for $0.6r_0<L\lessapprox 4.8r_0$, which shifts towards smaller energies in a quadratic way as a function of the length $L$, which is qualitatively in accordance to the behavior of eigenenergies in a quantum well that represent the position of the quasibound states. For $L>4.8r_0$, we start to observe two dips within the same plateau, which indicates the presence of two quasibound states within the same range of energy where the plateau takes place. To further explore effects of the Rashba SOC, we vary $\alpha=0.0\rightarrow 0.2\alpha_0$ with step $\Delta\alpha=0.02\alpha_0$ assuming $L>8r_0$, $E_Z=0.2\varepsilon_0$, and the cone-shape deformation with $\Delta r_0/r_0=0.3$. Such results are plotted in figure 11, which shows that there is no dip in the first plateau for $\alpha\leq 0.02\alpha_0$. Yet, dips appear for $\alpha>0.02\alpha_0$ resulting only from the Rashba SOC. Within others plateaus, dips coming from the radius expansion already appear for $\alpha=0$ and these dips interfere with dips descendant from Rashba SOC. Finally, we probe the conductance for different magnetic fields, considering $L>8r_0$, $E_Z=0\rightarrow0.2\varepsilon_0$ with step $\Delta E_z=0.02\varepsilon_0$, the square-shape deformation with $\Delta r_0/r_0=0.3$, and for two different values of Rashba constant $\alpha=0.2\alpha_0$ (figure 12) and $\alpha=0.8\alpha_0$ (figure 13). When the Rashba SOC energy $E_{R}=m^*\alpha^2/2\hbar^2$ is smaller than the Zeeman energy, the conductance is more affected by the magnetic field and more dips appear when $E_Z$ is increased, as shown in figure 12. On the other hand, figure 13 shows that dips are not very affected by the magnetic field because in this case the Rashba SOC dominates over the external magnetic field.

\section{Conclusions}
In this paper, we provided a method to numerically calculate the conductance considering an arbitrary 3D-spatial shaped potential and including the coupling between scattering channels. By employing this method, we were able to calculate the conductance for an NW containing structural deformations. We also showed the appearance of reentrant behavior without the Rashba SOC and magnetic field. Furthermore, we investigated the effects of structural deformations taking into account an external magnetic field. We found that these deformations affect the observation of the reentrant feature. Particularly, the compression of the NW radius can destroy the dip in the conductance. Here, we focused on structural deformation of the NW, but similar effects would occur in NWs under the presence of a localized potential. We believe the deformations can be experimentally achieved by employing nonuniform top-bottom gates in experiments such as those reported in Refs.~[\onlinecite{Heedt,Kammhuber}]. Also, III–V NWs grown in the [111] direction by the vapour-liquid-solid mechanism that exhibit planar stacking faults~\cite{Caroff,Algra} are a good platform to probe our findings.

\begin{acknowledgements}
The authors are grateful for financial support by  the Brazilian Agencies FAPESP, CNPq and CAPES. 
LKC thanks to the Brazilian Agencies FAPESP and CNPq (grant No 2019/09624-3 and No 311450/2019-9) for supporting this research. 
\end{acknowledgements}
\bibliography{refer}
\end{document}